\documentclass{IEEEtran}
\usepackage{cite}
\usepackage{amsmath,amssymb,amsfonts}
\usepackage{algorithmic}
\usepackage{textcomp}
\usepackage{bm}
\usepackage{algorithm}
\usepackage{balance}
\usepackage{graphicx,subfig,multicol,multirow,diagbox,booktabs,array}
\usepackage{color}
\usepackage{float}
\usepackage{stfloats}
\ifCLASSINFOpdf
\else
\fi
\begin{document}
\title{DOA Estimation Using Massive Receive MIMO: Basic Principle and Key Techniques}

\author{Jiangzhou Wang, Baihua Shi, Feng Shu, Qi Zhang, Di Wu, Qijuan Jie, \\Zhihong Zhuang, Siling Feng, and Yijin Zhang

\thanks{This work was supported in part by the National Natural Science Foundation of China (Nos.U22A2002, 61972093 and 62071234), Hainan Province Science and Technology Special Fund (ZDKJ2021022), and the Scientific Research Fund Project of Hainan University under Grant KYQD(ZR)-21008.}
\thanks{J. Wang is with the School of Engineering, University of Kent, Canterbury CT2 7NT, U.K. (e-mail: j.z.wang@kent.ac.uk).}
\thanks{B. Shi and Z. Zhuang are with the School of Electronic and Optical Engineering, Nanjing University of Science and Technology, Nanjing 210094, China.}
\thanks{F. Shu is with the School of Information and Communication Engineering, Hainan University, Haikou, 570228, China and also with the School of Electronic and Optical Engineering, Nanjing University of Science and Technology, Nanjing, 210094, China. (e-mail: shufeng0101@163.com)}
\thanks{Q. Zhang, D. Wu, Q. Jie and S. Feng is with the School of Information and Communication Engineering, Hainan University, Haikou, 570228, China.}

% <-this % stops a space
%\thanks{XXX, XXX, XXX and XXX are with the School of Electronic and Optical Engineering, Nanjing University of Science and Technology, 210094, CHINA. Email: {XXX, XXX, XXX, XXX@njust.edu.cn}.}
}% <-this % stops a space
\maketitle

\begin{abstract}
As massive multiple-input multiple-output (MIMO) becomes popular, direction of arrival (DOA) measurement has been made a real renaissance due to the high-resolution achieved. Thus, DOA estimation using massive MIMO is crucial. This paper is to present its basic principle and key techniques. It is anticipated that there are still many challenges in DOA estimation using massive receive MIMO, such as high circuit cost, high energy consumption and high complexity of the algorithm implementation. New researches and breakthroughs are given to deal with those problems. Then, a new architecture, hybrid analog and digital (HAD) massive receive MIMO with low-resolution analog to digital converters (ADCs), is presented to strike a good balance among circuit cost, complexity and performance. Then, a novel three-dimensional (3D) angle of arrival (AOA) localization method based on geometrical center is proposed to compute the position of a passive emitter using single base station equipped with an ultra-massive MIMO system. And, it can achieve the Cramer-Rao low bound (CRLB). Here, the performance loss is also analyzed to quantify the minimum number of bits. 
%DOA estimation will play a key role in lots of applications, such as directional modulation, beamforming tracking and alignment for 6G.
\end{abstract}


\IEEEpeerreviewmaketitle

\section{Introduction}
\begin{figure}
	\centering
	\includegraphics[width=0.4\textwidth]{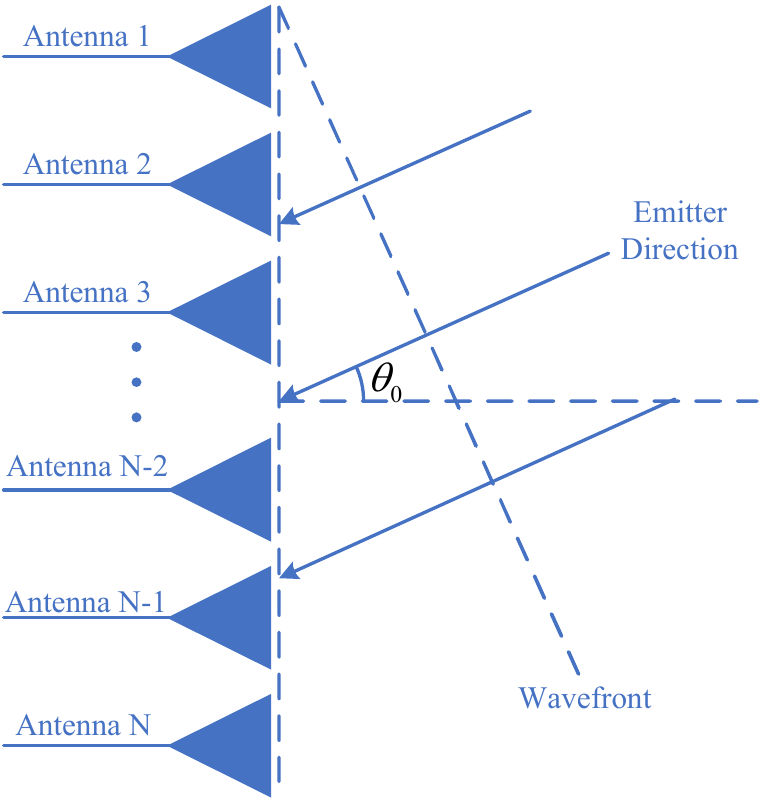}\\
	\caption{A uniformly-spaced linear array with M-elements receiving signals from a far-field emitter}
	\label{fig_DOA2ULA}
\end{figure}

Wireless direction-finding has emerged for a very long time since the wireless communication appeared. It supports various services such as localization and is expected to play an important role in beyond fifth generation (B5G) and sixth generation (6G) \cite{zhou2019cm}, which should be service-aware as firstly claimed in \cite{zhou2020Service}. Many basic concepts of direction-finding were built at that time \cite{keen1947}. Then, many advances in technology had been made since 1900s \cite{travers1966abstracts}. As opposed to the conventional active direction-finding techniques, direction of arrival (DOA) estimation is a passive direction-finding method. DOA estimation measures the direction by processing the phase difference of electromagnetic waves received by different antennas. DOA estimation has been applied in many fields, such as sonar, rescue, tracking of various objects, radio astronomy and other emergency assistance devices \cite{tuncer2009classical}. In the modern engineering applications, DOA estimation also plays an important role. For example, direction-finding is a new feature in Bluetooth 5.1 \cite{suryavanshi2019direction}. And, the current Bluetooth proximity system is able to measure the distance between different devices by making use of signal strength. Thus, by combining these techniques, the devices with Bluetooth 5.1 can figure out the location of a device among them \cite{suryavanshi2019direction}.
Consider a uniformly-spaced linear array (ULA) with M-antennas as shown in Fig. \ref{fig_DOA2ULA}. The narrow band signals from a far-field emitter will arrive at the array. It can be seen that the distances from the emitter to the different antennas are different. If we choose Antenna 1 as the reference point, the other antennas have their own distance differences to Antenna 1. Thus, the signals will arrive at the antennas at different times, which can result in the phase difference of the sampling data. The vector is composed by the phase difference of all antennas is called “array manifold”, which is written as $\mathbf{a}(\theta_0)$. And, the direction information is contained in the array manifold.

Based on that structure, some common methods are described to introduce basic concepts of DOA estimation. The first algorithm measures the DOA by using receive beamforming. Adjust the receive beamforming from $-\pi/2$  to $\pi/2$  and compute the output power of different angles. Signals of different antennas are added in phase only when the angle of receive beamforming matches the DOA. Thus, the estimated DOA is the angle corresponding to the largest value of output power as presented in . The second is called monopulse, often adopted for tracking the target in radar systems. Similar to the first, resorting to the output power of receive beamforming with different angle, this method calculates the power difference of two directions with little interval at all angles. For the symmetry of the output power to the DOA, the value of power difference will approach zero at the DOA. The methods above are both the linear search for the direction spectrum. And, when we need to estimate the directions of multi-signals, both the above methods are hard to be used. Because those methods fail to distinguish and measure the directions when the directions’ spacing of emitters are less than a beamwidth apart. Here, a spatial-spectrum-based algorithm is presented. It is a “super-resolution” estimation method and known as Multi-Signal Classification (MUSIC) algorithm \cite{schmidt1982music,Schmidt1986tap,weiss1994tsp}. 
Afterwards, a MUSIC-based search-free method was proposed in \cite{barabell1983ICASSP}, which is the famous root-MUISIC. In that algorithm, the spectrum search is replaced with root solving \cite{friedlander1993root}. Modifications of root-MUSIC have been proposed to improve the performance or reduce the computational complexity \cite{zoltowski1993beamspace,Pesavento2000tsp,yan2014tsp}. Another well-known search-free algorithm is estimation of signal parameters via rotational invariance technique (ESPRIT), which was proposed in \cite{Paulraj1986poi} and further improved in \cite{Roy1986ESPRIT,Roy1989ESPRIT,Ottersten1991tsp}. There is the same difference of array manifold for multi emitters in two subarrays. Thus, we can estimate the phase difference firstly. Then, the directions are measured by processing the phase differences. 

Massive multiple-input multiple-output (MIMO) has attracted much attention in recent years for its ten times increasement of achievable rate and energy efficiency. However, the beam width of main lobe becomes narrower and narrower as the number of antennas increases \cite{tse2005fundamentals}. Thus, the widely implementation of massive MIMO systems equipped with a large number of antennas demands the ultra-high precision of DOA. In addition, the huge increase of circuit cost and energy consumption in massive MIMO is also a challenge for wide applications. Many new array structures of massive MIMO systems have been proposed to handle that problem \cite{Singh2009tcom,Liang2016mixed,Ayach2014twc,sun2020twc}. New structures need new DOA algorithms and the corresponding performance analysis. More recently, according to the new requirement in wireless communications and other fields, some new progress of DOA estimation has been made in massive MIMO systems.

This paper provides a tutorial and presents research results in the DOA estimation using massive receive MIMO. The purpose of the DOA estimation is to measure angles of transmitters to receivers for the wireless communications and location. The implantation of the massive MIMO into the DOA estimation provides accurate channel estimation for wireless communications. From an overview of the progress made, we forecast that the DOA estimation using massive receive MIMO will play an important role in the sixth generation (6G), location and other future engineering applications. 

%The remainder of this paper is organized as follows. Passive target detection is given in Section 2. Section 3 focuses on the estimation methods and key technologies.  The performance analysis for the low-resolution ADC structure, mixed-ADC structure and HAD structure is included in Section 4. In addition, we proposed a new AOA intersection method in Section 5. Finally, conclusions are summarized in Section 6.

\section{Passive emitter detection}
Passive emitter detection is a key part for DOA estimation. Only when an emitter is detected, a DOA estimation is triggered. In this section, we will give a brief introduction of the passive target detection.
\cite{Kelly1986taes} gave general problems of signal detection and focused on the likelihood ratio test rule. In \cite{Orlando2010tsp,Niu2012taes}, authors proposed the detection of target with MIMO radar and explored many target detection algorithms. By adopting eigenvalue decomposition, the independent observation components were extracted from M orthogonal signals and weighted together. In \cite{douganccay2004passive}, the linear least squares method was used to solve the radiation source location. Therefore, target detection method was preferred for more practical applications. Non-cooperative passive bistatic radar has become a research hotspot because of its low cost, anti-interception, anti-stealth and other merits \cite{hack2014tsp}. The passive radar is not equipped with a dedicated transmitter, but uses an existing wireless source as an illuminator of opportunity to detect and track the target \cite{Degurse2014taes}. In \cite{Karthik2018spl}, some reduced-rank space-time adaptive processing (STAP) algorithms were proposed. Compared with traditional STAP algorithms, less sample data is required in these methods. However, they are sub-optimal algorithms due to the trade-off between performance and computational complexity. \cite{Sirianunpiboon2014rc} proposed a method to detect the existence of the transmitted signal by using the rank of the transmitted signal. Furthermore, they extended this method to the problem of MIMO radar signal characterization, which achieved a better performance. 

\begin{figure}
	\centering
	\includegraphics[width=0.48\textwidth]{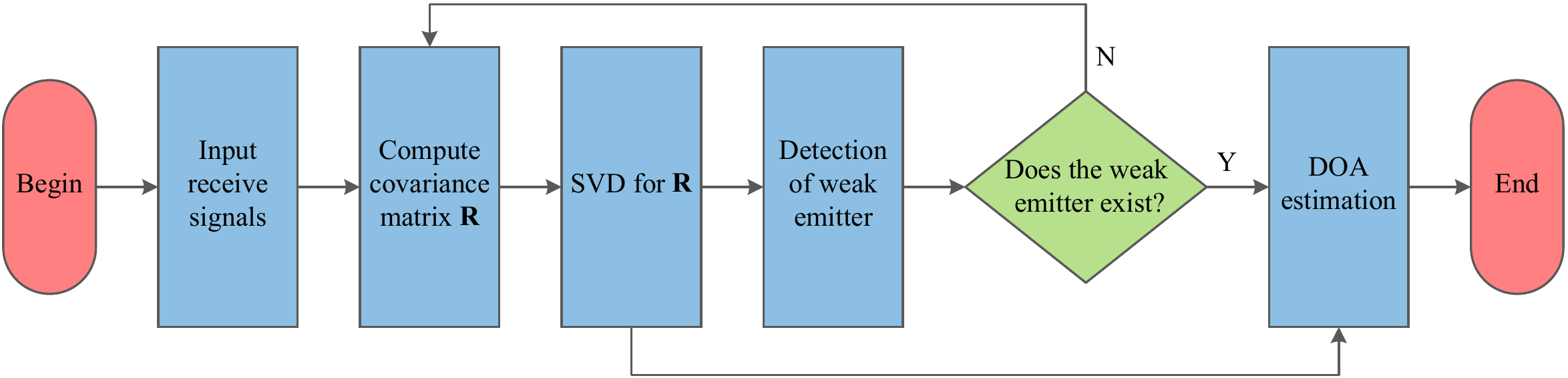}\\
	\caption{Block diagram of the proposed detection model}
	\label{fig_detect}
\end{figure}

\begin{figure}
	\centering
	\includegraphics[width=0.48\textwidth]{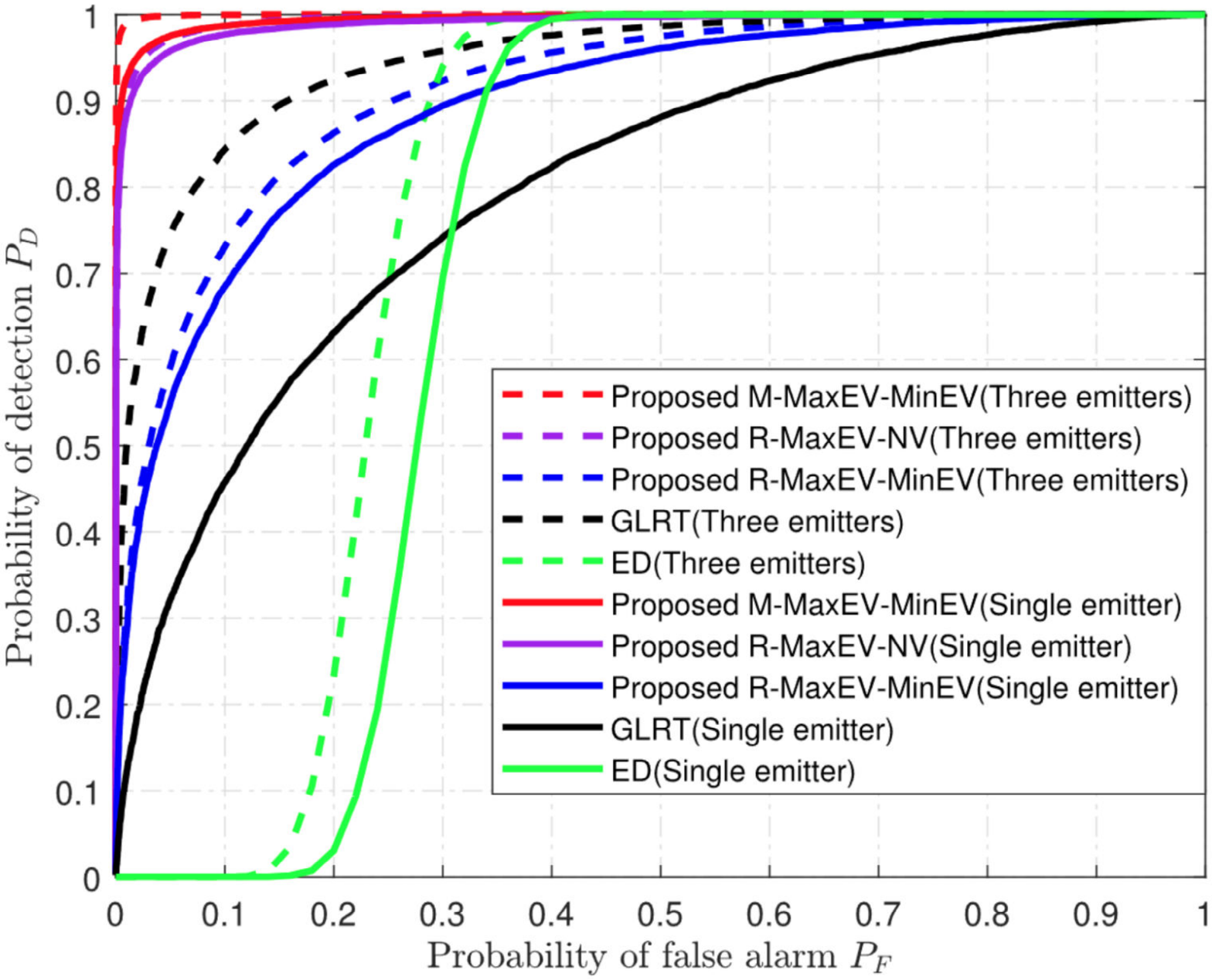}\\
	\caption{Probability of detection with proposed methods in \cite{Jie2022wcl} and other methods}
	\label{fig_detect_simu}
\end{figure}
Motivated by the idea of target detection from radar, \cite{Jie2022wcl} considered a new EVD-based passive target detection model. In Fig. \ref{fig_detect}, a block diagram of DOA measurements with the ability of detecting a weak emitter is sketched. Here, the sampling covariance of receive signal vector is computed, and its EVD is performed to extract all its eigenvalues. Three methods were proposed. The first method is defined as the ratio of the largest eigenvalue to the smallest one while the second one is the ratio of the largest eigenvalue to the estimated noise variance.  The statistic test is given as the arithmetic mean of the largest and smallest eigenvalues. From simulation results, shown in Fig. \ref{fig_detect_simu}, the proposed three methods perform better than the traditional generalized likelihood ratio test method with a given false alarm probability.

\section{Fundamentals of DOA Estimation}
In this section, fundamentals of DOA estimation are presented, including some classic methods, performance metrics.

\subsection{Comparisons of Classic Methods}

The conventional beamforming method is to align the array response from $-90^{\circ}$ to $90^{\circ}$. The output power spectrum will have peaks at correct directions.
One of disadvantages is that this method could not estimate multiple emitters with close spacing.

\begin{figure*}
	\centering
	\includegraphics[width=0.7\textwidth]{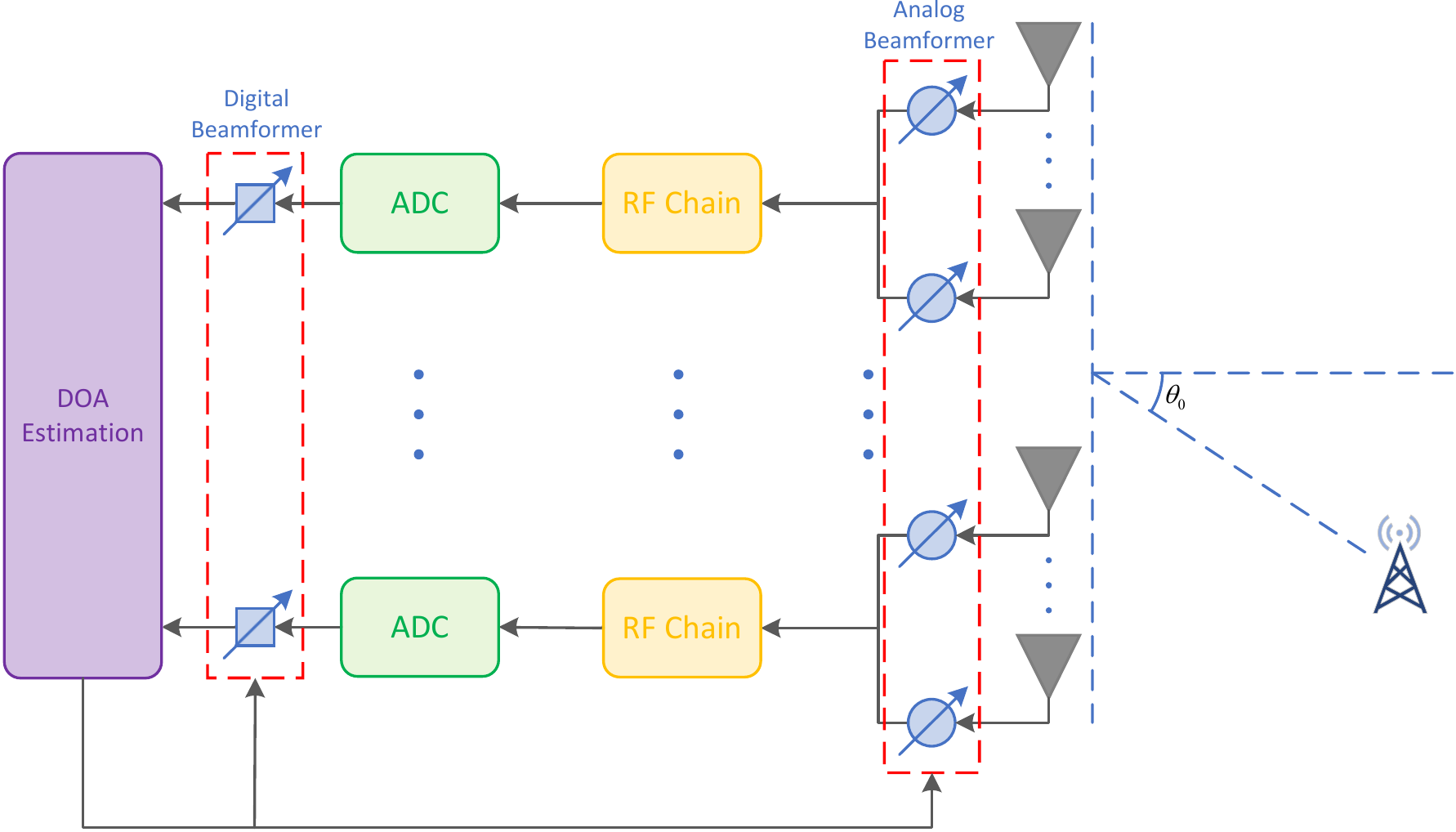}\\
	\caption{System model for the sub-connected HAD architecture}
	\label{fig_structure_HAD}
\end{figure*}
\begin{figure}
	\centering
	\includegraphics[width=0.49\textwidth]{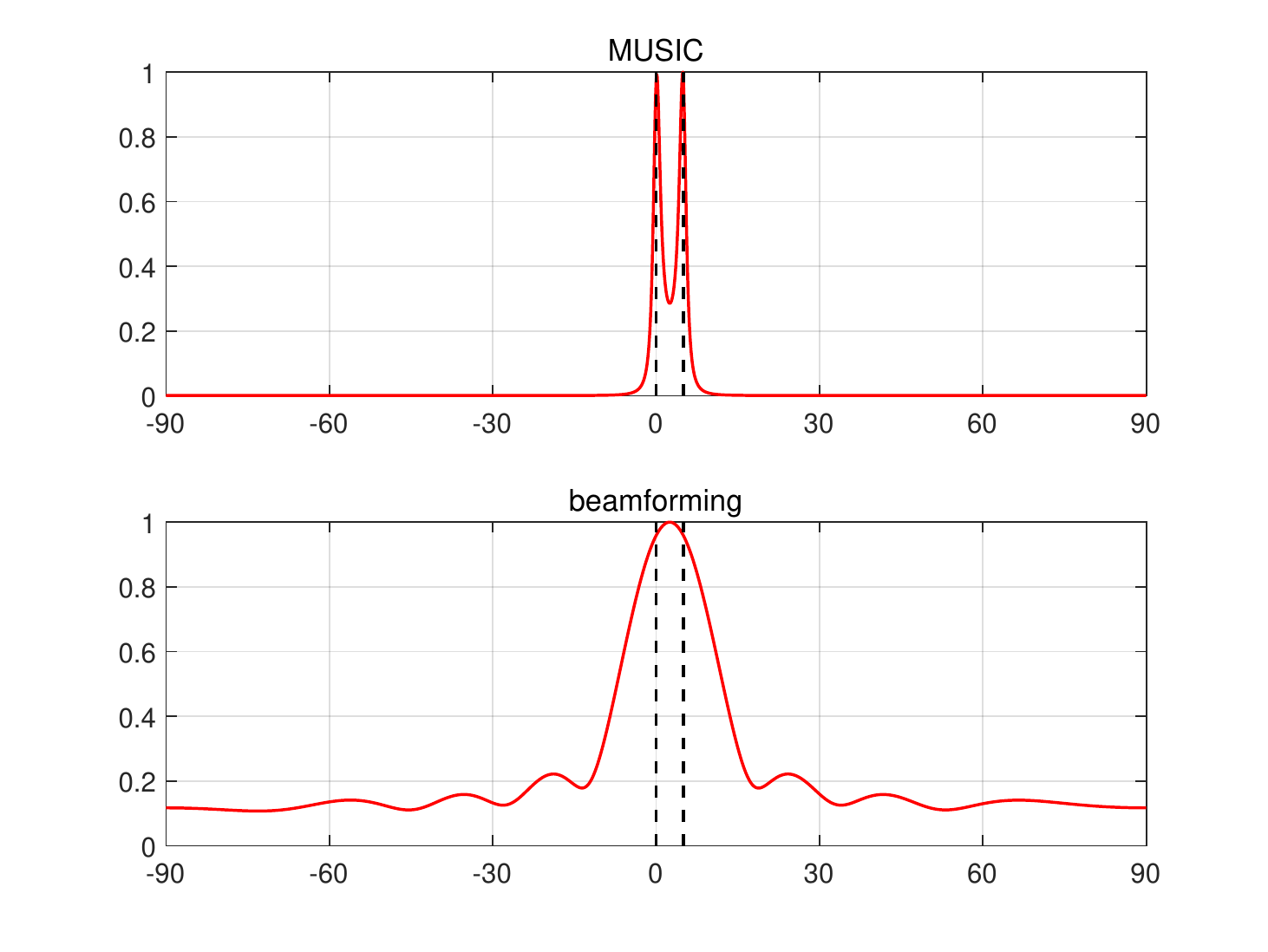}\\
	\caption{The spectrum of the beamforming and MUSIC for two uncorrelated signals at $0^{\circ}$ and $5^{\circ}$}
	\label{fig_music}
\end{figure}
Subspace based methods could fix this problem, such as Capon, ESPRIT, and MUSIC.
MUSIC method construct the covariance matrix first.
Then, eigenvalue decomposition is conducted on the covariance matrix.
Since the signal subspace and noise subspace are orthogonal, MUSIC has higher resolution. As presented in \ref{fig_music}, when two angles are set as $0^{\circ}$ and $5^{\circ}$, beamforming fail to separate two angles but MUSIC method is able to do that.

\subsection{Cramer-Rao Lower Bound}
Cramer-Rao lower bound (CRLB) is the lower bound of minimum variance for unbiased estimation. This is adopted to evaluate the estimator accuracy in DOA estimation. In theory, accuracy of the proposed methods could approach but could not exceed the CRLB. In addition, the closed-form expression of CRLB is only existed for the single emitter. Thus, most works were focused on the single emitter scenario.

\subsection{Resolution}
In most applications, the system needs to estimate multiple emitters at same time. Thus, the ability of distinguishing closely spaced emitters is also an important metric in DOA estimation. Accuracy depends on the estimation and resolution depends on detection and decision \cite{kay1993fundamentals}.

For conventional methods, the resolution ie equal to the beamwith. In other words, when the space of two directions is less than the beamwith, the methods can not separate them. For subspace based methods, the resolution is related to the SNR, which is given by
\begin{equation}
	\bigtriangleup r \approx \frac{BW}{SNR^{\frac{1}{4}}} \nonumber
\end{equation}
where $\bigtriangleup r$ and $BW$ denotes the resolution and beamwith, respectively.
\begin{figure}
	\centering
	\includegraphics[width=0.49\textwidth]{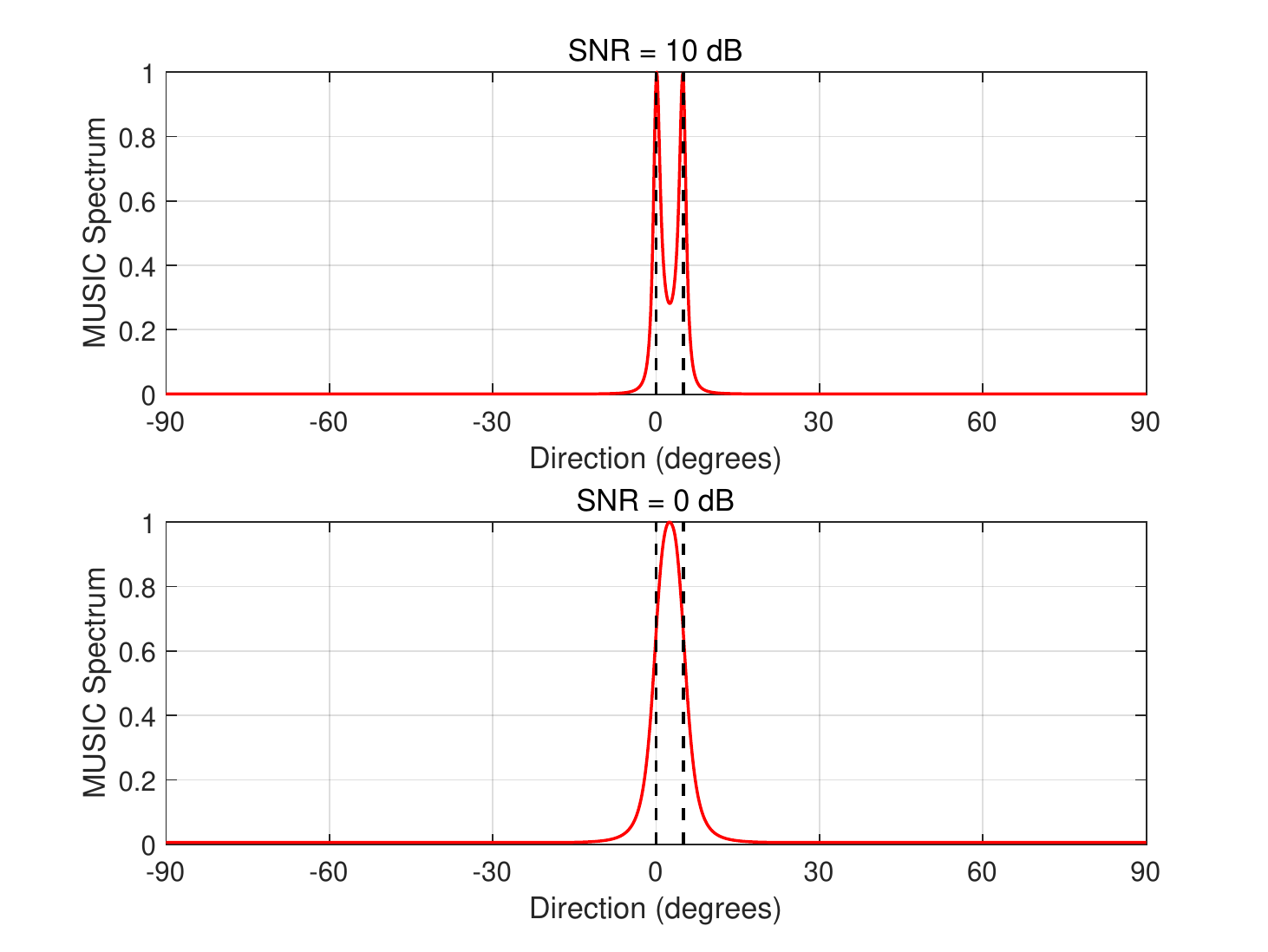}\\
	\caption{The spectrum of the  MUSIC for two uncorrelated signals at $0^{\circ}$ and $5^{\circ}$ with different SNR}
	\label{fig_musicDiffSNR}
\end{figure}

As shown in Fig. \ref{fig_musicDiffSNR}, when SNR=10 dB, MUSIC method could distinguish two emitters. However, as SNR was decreased to 0 dB, two directions will merge into one peak.

\section{Advanced DOA estimation methods and localization}
In this section, DOA estimation in the HAD structure is presented to illustrate the key technology for the DOA estimation using massive receive MIMO. And, some researches on other antenna structures are given as well. 

\subsection{DOA estimation for the hybrid architecture and angle ambiguity elimination}
The key challenges of the DOA estimation using massive receive MIMO is how to reduce circuit cost, power consumption and computational complexity while maintaining a high accuracy. The implementation of hybrid analog and digital (HAD) structure is a promising solution \cite{Han2015cm}. This architecture was firstly proposed in \cite{Zhang2005tsp,Sudarshan2006twc} and has been investigated in massive MIMO systems in recent years \cite{Bogale2013gcc,Sohrabi2016jstsp}.

A $K\times M$ sub-connected HAD structure is shown in Fig. \ref{fig_structure_HAD}, where the array has $K$ subarrays, and there are $M$ antennas in each subarray. Each subarray is connected to a radio frequency (RF) chain. In this architecture, the analog beamforming is conducted by adjusting the phases of analog phase shifters. After going through the RF chains, the frequency band signals are converted into the baseband signals. Then, the analog signals are quantified to the digital signals by analog-to-digital converter (ADCs). The digital beamforming can control the phases and digital weights for the signals. Obviously, compared with the conventional ULA requiring $N(N=KM)$  RF chains, the HAD architecture just needs $K$  RF chains, which saves enormous hardware cost and energy consumption. 

Since the antenna spacing is half wavelength, the interval between the centers of two adjacent subarray is $M\lambda/2$ . Thus, if we adopt the conventional estimation methods directly, like MUSIC and ESPRIT, the phase ambiguity will arise. This is the challenge that has to be addressed for the DOA estimation in the HAD structure \cite{Zhang2016cl}. 

\begin{figure}
	\centering
	\includegraphics[width=0.48\textwidth]{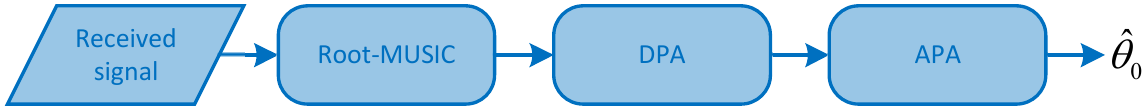}\\
	\caption{Outline flowchart for the root-MUSIC-HDAPA method}
	\label{fig_HADrootMUSIC_flowchart}
\end{figure}

In \cite{shu2018tcom}, two phase alignment (PA) methods and one root-MUSIC-based method were proposed. There are two steps in the first PA method: analog phase alignment (APA) in the first step and digital phase alignment (DPA) in the second step. Search for the maximum power and the estimated direction is calculated from the corresponding phase. The second method is to reverse the order of APA and DPA in the first method. The above two methods have no problem with phase ambiguity. However, both methods need linear search. Shortening the search stepsize can increase the accuracy, but the computational complexity and processing time also increase. The proposed root-MUSIC-based method, called root-MUSIC-HDAPA, measure the direction by three steps as shown in Fig. \ref{fig_HADrootMUSIC_flowchart}. Adopt the root-MUSIC at first. Then, the DPA is used to elect the correct phase. Finally, the APA is employed to eliminate the phase ambiguity. This is a search-free method. Simulation results reveal that root-MUSIC-HDAPA and HDAPA could achieve the hybrid CRLB with very low computational complexity. In addition, if high complexity is acceptable, HADPA and APA can be employed to reach the full-digital CRLB.

\begin{figure}
	\centering
	\includegraphics[width=0.48\textwidth]{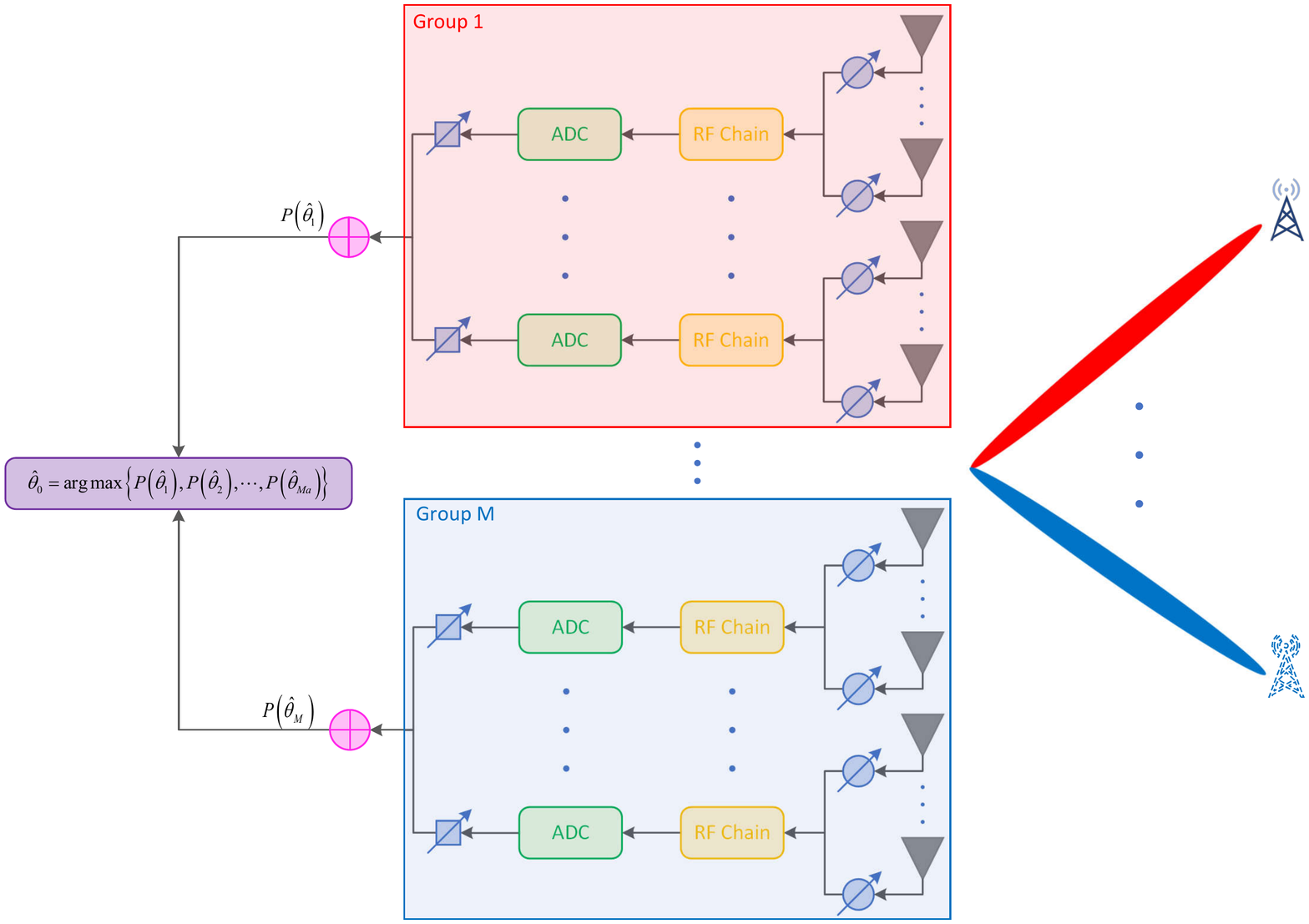}\\
	\caption{ Proposed structure for the fast phase ambiguity elimination}
	\label{fig_reAmb}
\end{figure}

However, that method still needs additional $M$  time blocks for APA. To handle this problem, an improved fast ambiguous elimination method was proposed in \cite{shi2022scis}. This method modifies the part of APA. Divide the subarrays into $M$  groups. And, subarrays in each group align their analog beamforming to one candidate angle in a time block, as shown in Fig. 5. Then, the phase ambiguity is eliminated. The most important advantage of this improved method is that it just requires two time blocks to estimate the DOA , as compared to the $M+1$  time blocks in the original method in \cite{shu2018tcom}. The computational complexity also decreases from  $O(K^2L+8(K-1)+L((2K-2)K+MN))$ to $O(K^2L+8(K-1)+L((2K-2)K+N))$, where  $L$ is the number of snapshots. However, because fewer antennas are used to align the phase, the performance of the improved method is worse than the original method. This method also claims for $K\geq M$. 

To reduce the vast computational complexity for the DOA estimation in massive MIMO systems, the authors firstly introduced the deep learning into DOA estimation in \cite{huang2018tvt}, which significantly decreases the complexity. By resorting to the deep neural network (DNN), two high-resolution schemes were proposed to conduct DOA estimation and channel estimation. Inspired by \cite{shu2018tcom}, authors proposed a ESPRIT-based method for the HAD structure in \cite{zhuang2020scis}. To improve the precision of estimated value, a machine-learning framework was considered. It is worth noting that the proposed phase ambiguity elimination method in \cite{shi2022scis} is also suitable for the ESPRIT-based method in \cite{zhuang2020scis}.

In \cite{li2020cl}, a special case for the HAD structure was investigated. All Antennas are connected to one RF chain, which means $M=N$.  A set of angles, $\{\theta^1,\theta^2,\cdots,\theta^Q\}$, is predetermined. Then, the analog shifters are aligning the phases to these angles in turn. Through analysis and comparison, the covariance matrix can be reconstructed by the average power for the different angles. And, covariance matrix reconstruction almost has no extra calculation because most operations can be pre-calculated off-line. This antenna structure is very simple and cheap. Complexity of the proposed method is also very low. However, similar to other beam-sweeping-based methods, the shortcoming of this method is that the accuracy is affected by the $Q$. In addition, it needs  $Q$ time blocks to sample data, which consume too much processing time.

To reduce the huge computational complexity in massive MIMO systems, \cite{chenDOA2022wcl} proposed three subspace based methods. Two of these methods divide the array into some subarray. The final measured value is achieved by combining angles estimated by subarrays. Furthermore, the third method is a successive convex approximation (SCA) based estimator and the initial value is generated by a subarray. These methods have much lower computational complexities and are able to achieve the CRLB.

\subsection{DOA estimation for 2D array}
In massive MIMO systems, to make the base stations (BSs) accommodate antennas, two-dimensional (2D) arrays are preferred to be employed. In 2D massive MIMO systems, there are two directions needed to be measured: elevation angle and azimuth angle. Some pioneering works directly expanded the 1D methods to 2D arrays, giving birth to the 2D-MUSIC \cite{gao2009ic}, 2D-ESPRIT \cite{Zoltowski1996tsp} and 2D unitary matrix pencil method (2D UMP) \cite{Yilmazer2006apc} and so on. However, the extremely high hardware cost and computational complexity in massive MIMO systems make it hard to be applied. And, the unavoidable bad data caused by massive MIMO systems has a significantly effect on the EVD \cite{cheng2015tsp,Larsson2014cm}. Thus, a completely new algorithm for the 2D array in the massive MIMO system is needed.

In \cite{cheng2015tsp}, a novel iterative algorithm for the massive MIMO systems with a 2D array was proposed. This method is based on the variational Bayesian framework and is able to estimate the direction with no access to the signal path number, noise power, path gain correlations and bad data statistics. Furthermore, considering that tensors are usually used in the 2D MIMO \cite{Kolda2009tensor}, this method is based on the tensor representation and operation. A sub-connected HAD structure with 2D massive MIMO was considered in \cite{wu2018twc}. A new analog beamforming was designed to eliminate the phase ambiguity. In this method, the analog shifters’ values of different subarrays are all different. Through derivation, the angle could be estimated by adopting cross-correlations of the adjacent subarrays and the phase ambiguity is eliminated by using the inverse discrete Fourier transform (IDFT) on the receive signals. However, this method can only be used to estimate the single emitter. In \cite{fan2018twc}, authors investigated a multiusers millimeter-wave (mm-wave) massive MIMO system with the full-connected HAD structure. By employing 2D DFT and optimal angle rotation, the angles of multiusers are estimated. Furthermore, a channel gain estimation algorithm was proposed.

\begin{figure*}[t]
	\centering
	\includegraphics[width=0.8\textwidth]{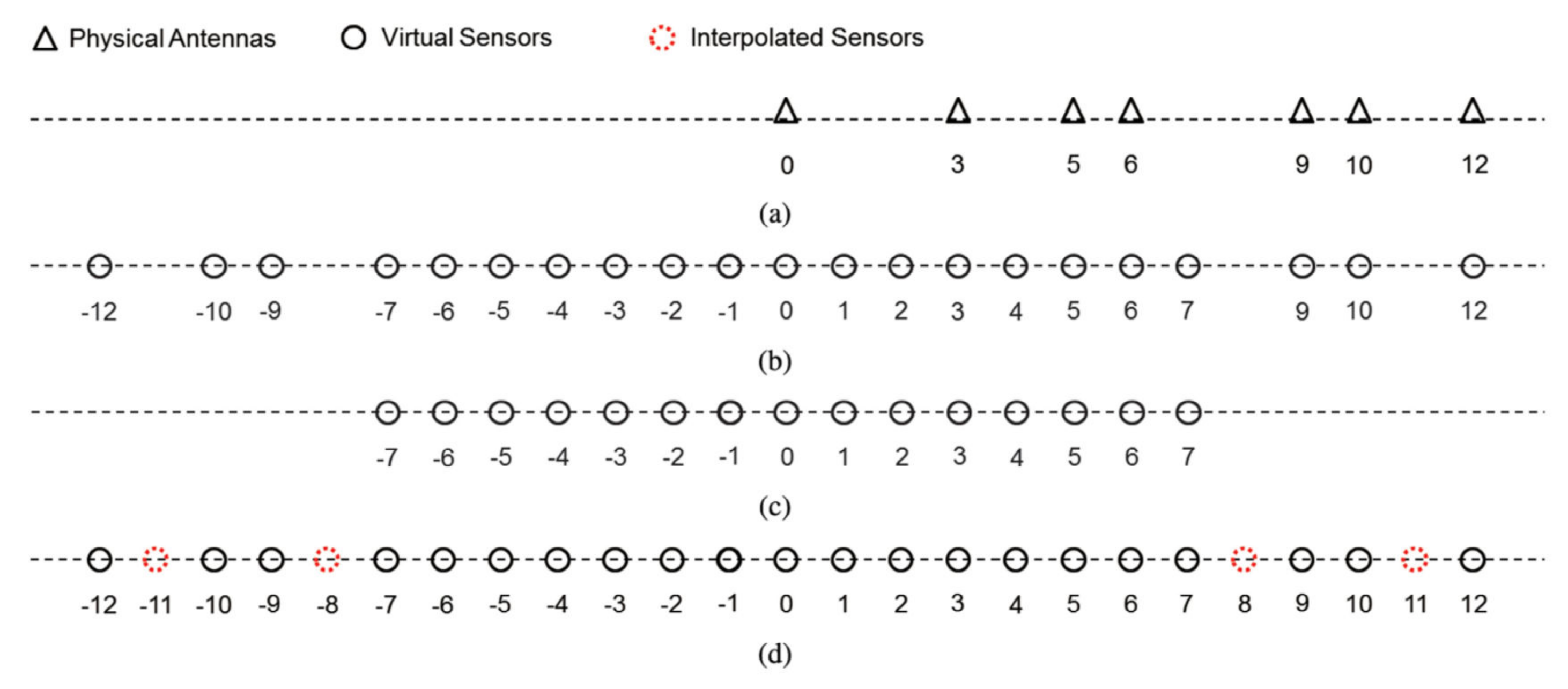}\\
	\caption{Illustration of various array representations in co-prime array. (a) Co-prime array. (b) virtual array derived from co-prime array. (c) Contiguous part of the virtual array. (d) Interpolated virtual array.}
	\label{fig_coprime}
\end{figure*}
\subsection{DOA estimation for uniform circular array and non-uniform array}
Compared with the HAD structure with ULA or 2D array, the hybrid massive MIMO system with the uniform circular array (UCA) has no problems with the phase ambiguity. Although some DOA estimation methods for the UCA has been developed for many years \cite{Mathews1994tsp,Jackson2015tap}, all of them cannot be applied in the HAD structure with UCA without modification. However, resulting from its complex structure, it is difficult to design the algorithm and analyze the performance. To our knowledge, the DOA estimation for this architecture is still an open problem. A deep-learning method for that architecture was proposed in \cite{hu2020wcl}. This method is based on the idea that convert the DOA estimation to the function fitting. Since the training for the deep feedforward networks is offline, the complexity of the real-time estimation will not be very high. 

\begin{figure}
	\centering
	\includegraphics[width=0.48\textwidth]{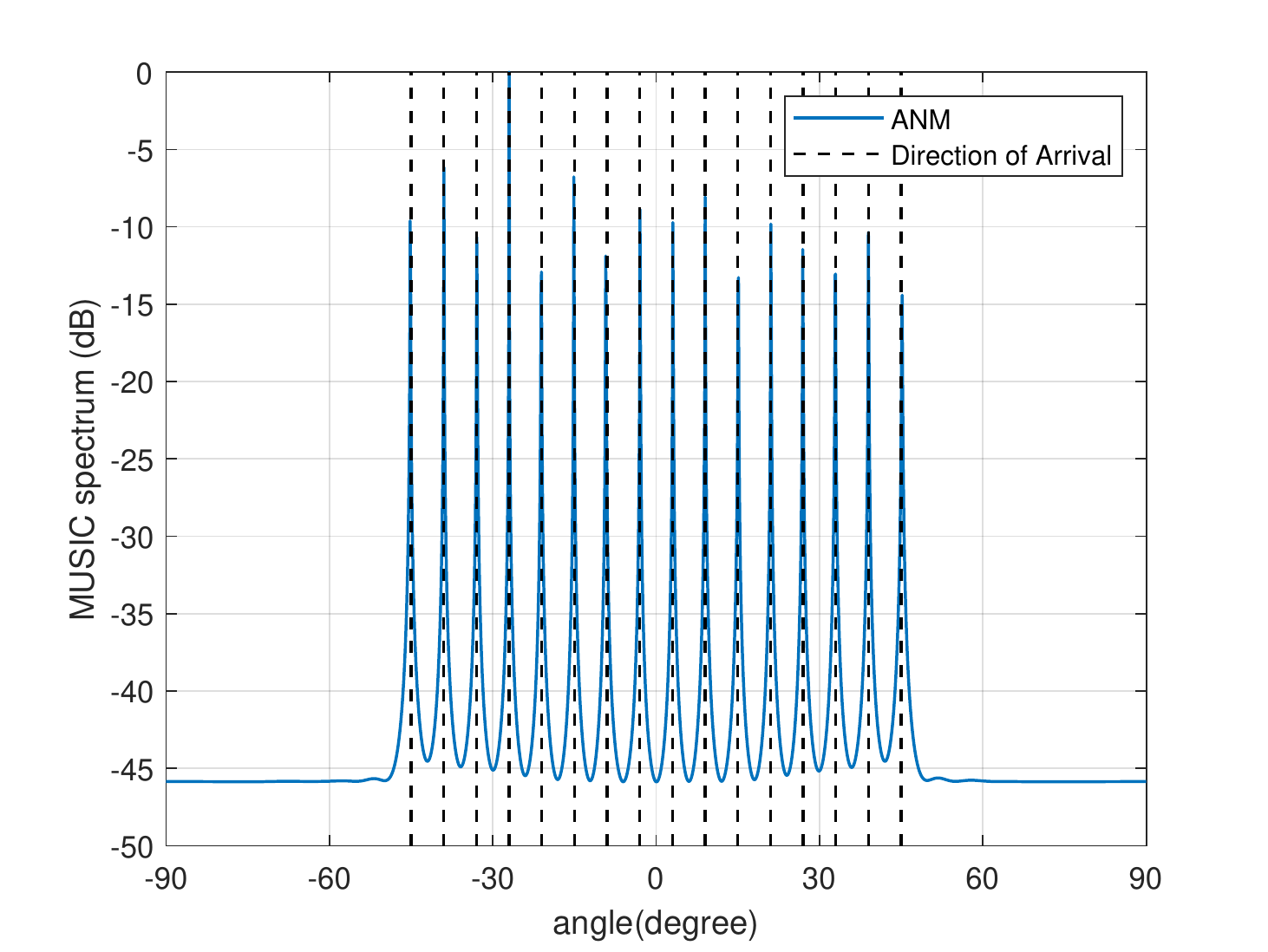}\\
	\caption{The MUSIC spectrum of the co-prime array with 10 antennas}
	\label{fig_coprime_simu}
\end{figure}
The non-uniform array (NUA) has better performance than the ULA with the same number of antennas and worse than the ULA with the same length \cite{tuncer2007direction}. Many works have been focused on some special array architectures, such as the nested array \cite{pal2010tsp}, co-prime array \cite{Vaidyanathan2011tsp,zhou2018spl,zhou2018tsp},  and sparse array \cite{wang2019spl}. Among them, co-prime array has attracted the most attention since it could increase degrees-of-freedom (DoFs). 
As illustrated in Fig. \ref{fig_coprime}, when the co-prime array is shown as Fig. \ref{fig_coprime}a, the virtual array can be obtained by vectorizing the covariance matrix, as shown as Fig. \ref{fig_coprime}b. However, there are still some holes in the virtual array. And, the contiguous part of the virtual array is presented in Fig. \ref{fig_coprime}c. To deal with this problem, atomic norm was employed to fill up theses holes in many existed works \cite{Vaidyanathan2011tsp,zhou2018spl,zhou2018tsp}. The interpolated virtual array is shown in \ref{fig_coprime}d. Then, MUSIC can be adopted. For example, the MUSIC spectrum of a co-prime array with 10 antennas is presented in Fig. \ref{fig_coprime_simu}. The DoFs of a ULA with 10 antennas is 10. And, the co-prime  array could separate these 16 directions. This means that the DoFs is much higher than a ULA with 10 antennas.
In \cite{Wagner2021tsp}, grid-less methods for arbitrary array geometries were proposed. In order to extend root-MUSIC to the NUA, authors introduced the irregular Vandermonde matrix and irregular toeplitz matrix. Moreover, alternating directions method of multipliers is used to solve the non-convex rank minimization problem in the NUA \cite{body2004convex,boyd2011distributed}. 

\section{CRLB, performance loss and energy efficiency}
Adopting low-resolution ADCs is another promising solution to reduce the hardware cost and energy consumption. The massive MIMO system with low-resolution ADCs has attracted much attention in recent years. Many performance analyses on the spectral efficiency have been investigated \cite{fan2015cl,zhang2016cllow,dong2018cl,dong2020tvt}. In \cite{bar2002taes}, a reconstruction  for the covariance matrix of unquantized signals was proposed. Then, some similar methods for the sparse arrays were presented in \cite{liu2017icassp,chen2018access}. The compressive sensing was adopted to handle the problems of one-bit DOA estimation in \cite{Christoph2015spawc,yu2016spl}. Different from the above methods, authors proved that the quantized one-bit signals can be used to estimate DOA by MUSIC without extra modification. Afterwards, the CRLB and performance loss were investigated in \cite{shi2022sj}. With the help of the additive quantization noise model (AQNM) in \cite{Orhan2015ita}, the nonlinear quantization of the low-resolution ADCs is transferred to a  linear gain with the additive quantization noise. Then, with the help of CRLB, we defined a performance loss factor to assess the performance. The CRLB is a lower bound on the variance of unbiased estimators. It is an important benchmark for the estimation methods.

\begin{figure}
	\centering
	\includegraphics[width=0.48\textwidth]{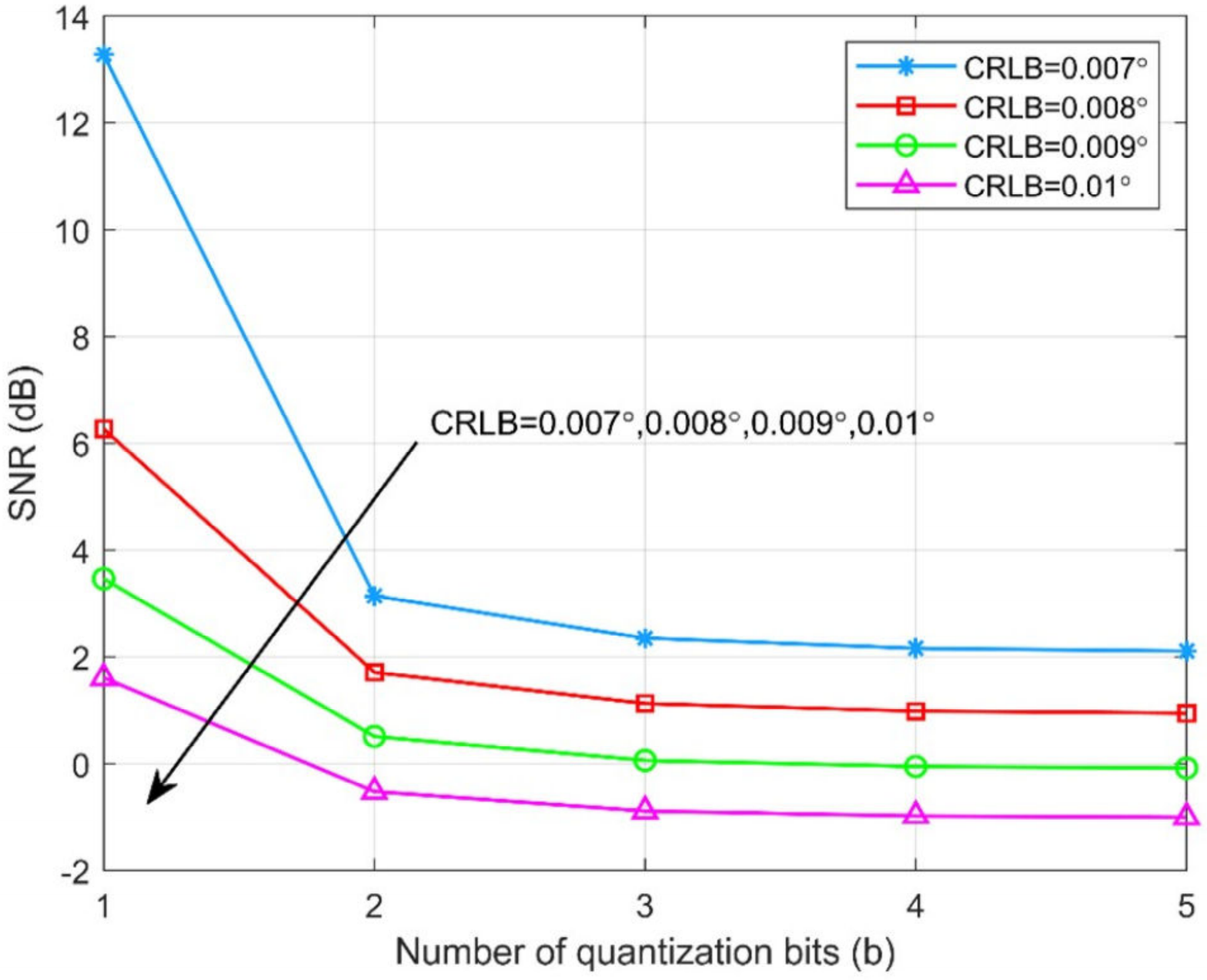}\\
	\caption{Trade-off between the quantization bit and SNR}
	\label{fig_lowADC_snrb}
\end{figure}

The trade-off between the quantization bit and SNR for different CRLB is illustrated in Fig. \ref{fig_lowADC_snrb}. As can be seen, all curves decrease as the quantization bit increases. Obviously, if we just pursuit a constant CRLB rather than the low performance loss, the quantization bit could be reduced as the SNR increases. An insightful observation is that the performance has negligible improvement when the quantization bit is increased from 4 to 5. However, the power consumption and circuit cost increase significantly. Thus, we had better choose 1-4 bits of ADCs’ resolution.

However, there are still some challenges in the low-resolution structure, such as time-frequency synchronization, achievable rate and so on \cite{Liang2016mixed}. Thus, some researches on the achievable rate have been done \cite{zhang2017mixed,zhang2019tcom}. The DOA estimation for the mixed-ADC architecture was considered by us in \cite{shi2022wcl}. We also proved that the quantized signals are able to be utilized in the MUSIC method. Referring to the energy efficiency in \cite{zhang2017mixed}, an energy efficiency factor for the DOA estimation was introduced, which is given by

\begin{equation}
	\eta=\frac{CRLB^{-\frac{1}{2}}}{P_{total}}~1/degree/W \nonumber
\end{equation}
where $P_{total}$  is the total power consumption of the antenna array. In \cite{shi2023ojcs}, we extended the DOA estimation with mixed-ADCs to the HAD structure. The closed-form expression of the corresponding CRLB was derived. We also investigated the performance loss and energy efficiency. The HAD structure with mixed-ADCs can be regarded as a more general form of the low-resolution ADC structure and mixed-ADC structure with the ULA. When we set 1 as the number of antennas in each subarray, the HAD structure will degenerated into the ULA.

\begin{figure}
	\centering
	\includegraphics[width=0.48\textwidth]{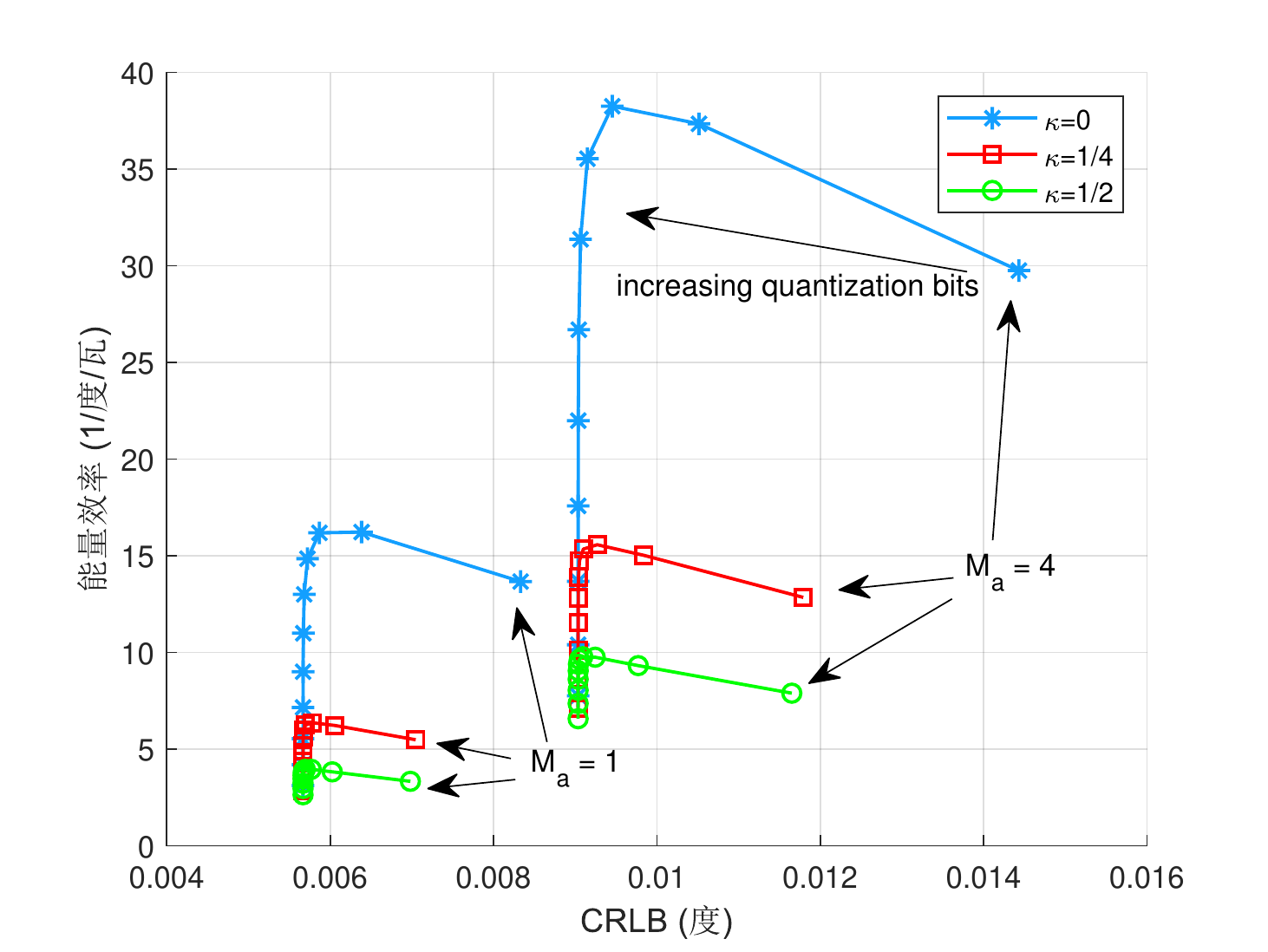}\\
	\caption{Trade-off between the CRLB and the energy efficiency}
	\label{fig_EE_CRLB}
\end{figure}
In Fig. \ref{fig_EE_CRLB}, the trade-off between the CRLB and energy efficiency with different proportions of the high-resolution ADCs is showed. $M_a$  is the number of antennas in each subarray. The curves of the full digital structure with mixed-ADCs ($M_a=1$) are plotted as a comparison. Among them, the blue curve ($M_0=0$) could represent the low-resolution ADC structure in \cite{shi2022sj}. It can be seen that HAD architecture with mixed-ADCs has much higher energy efficiency. And, adopting pure low-resolution ADCs can achieve the best energy efficiency. However, considering the poor performance of DOA estimation in Fig. \ref{fig_lowADC_snrb} and low achievable rate in \cite{zhang2017mixed}, it is difficult to adopt the pure low-resolution ADCs structure in the practical wireless communication systems. Moreover, we can conclude that the energy efficiency increases as the proportion of the high-resolution ADCs decreases. When we increase the number of quantization bits from 1 to 4, the CRLB will decrease quickly. While the performance almost has no improvement when quantization bits range from 4 to 12 bits. In addition, increasing the quantization bit of low-resolution ADCs in the HAD structure has more performance improvement than that in the full digital structure. 

Interested in the supreme energy efficiency of the HAD structure with pure low-resolution ADCs, we propose a HAD structure with low-resolution ADCs. To avoid the disadvantages discussed before, we could adopt this structure into the applications with no requirement to send information. For example, the passive radar using massive MIMO always works all day and never send any signals. The supreme energy efficiency is what it needs and the low achievable rate is what it not cares. Thus, the proposed structure is very suitable to that. 

\section{Proposed localization method via AOA intersection}
DOA estimation using massive MIMO systems could achieve an ultra-high precision of angles, which could pave the way to the angle of arrival (AOA) localization. AOA localization  points out the position of the emitter by processing the arrival angles received by sensors in the wireless sensor network (WSN) \cite{Torrieri1984taes}. Different from the time of arrival (TOA) and time difference of arrival (TDOA), AOA localization does not need the synchronization of receivers \cite{zhang2013twc}. Some iterative solutions and closed-form methods for the 2D AOA have been studied many years ago \cite{Torrieri1984taes,Lingren1978taes,shao2014tsp,wang2012tsp}. Due to the fact that the azimuth angle and the elevation angle are joint but nonlinear with the position, 3D AOA localization is more challenging. In recent years, many attentions have been attracted on 3D AOA \cite{wang2015twc,wang2018twc}. This passive localization could be applied in many 6G applications, like BS positioning, UAV localization \cite{li2021cl}, tracking \cite{meng2013pccc} and so on.

\begin{figure}
	\centering
	\includegraphics[width=0.47\textwidth]{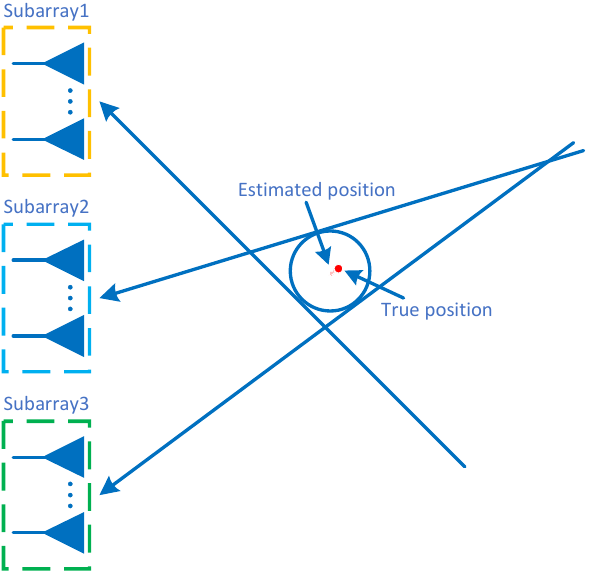}\\
	\caption{Diagram of AOA intersection localization via the concept of on geometrical center with 3 subarrays independently measure the DOA of emitter and the corresponding DOA lines intersect to form a triangle whose center being the position of emitter}
	\label{fig_AOA_2D}
\end{figure}
In this paper, we propose a 3D AOA method based on geometrical center to estimate the position of the passive emitter using a single BS equipped with an ultra-massive MIMO. In ultra-massive MIMO systems, the number of antennas and the size of the array are all huge. Divide the array into some subarrays with relatively long spacing. These subarrays can be regarded as sensors in the conventional WSN. Each subarray computes the DOA by itself. Then, the position is the geometric center of the intersection area of these estimated DOAs as shown in Fig. \ref{fig_reAmb}.

As plotted in Fig. \ref{fig_AOA_2D}, a 2D diagram of our proposed method is shown. For an AOA system with 3 subarrays, the point with the minimum distance sum is the center of inscribed circle for the DOAs. Extending this method to the 3D scene, there are two steps in our method: building multiple planes and minimizing the distances sum from the estimated point to all planes. The planes are structured by the directions of azimuth angles and the elevation angles measured by all subarrays. The minimizing sum of distances problem can be written as the form of
\begin{equation}
	\min_{\mathbf{u}=[x,y,z]^T}~~~\|\mathbf{A}\mathbf{u}-\mathbf{b}\|_1
\end{equation}
which is the sum of absolute residuals approximation. That also can be expressed as an linear programming (LP) \cite{body2004convex}. This problem itself is convex. To obtain the solution, we can resort to some standard convex optimization tools (e.g., CVX).
\begin{figure}
	\centering
	\includegraphics[width=0.49\textwidth]{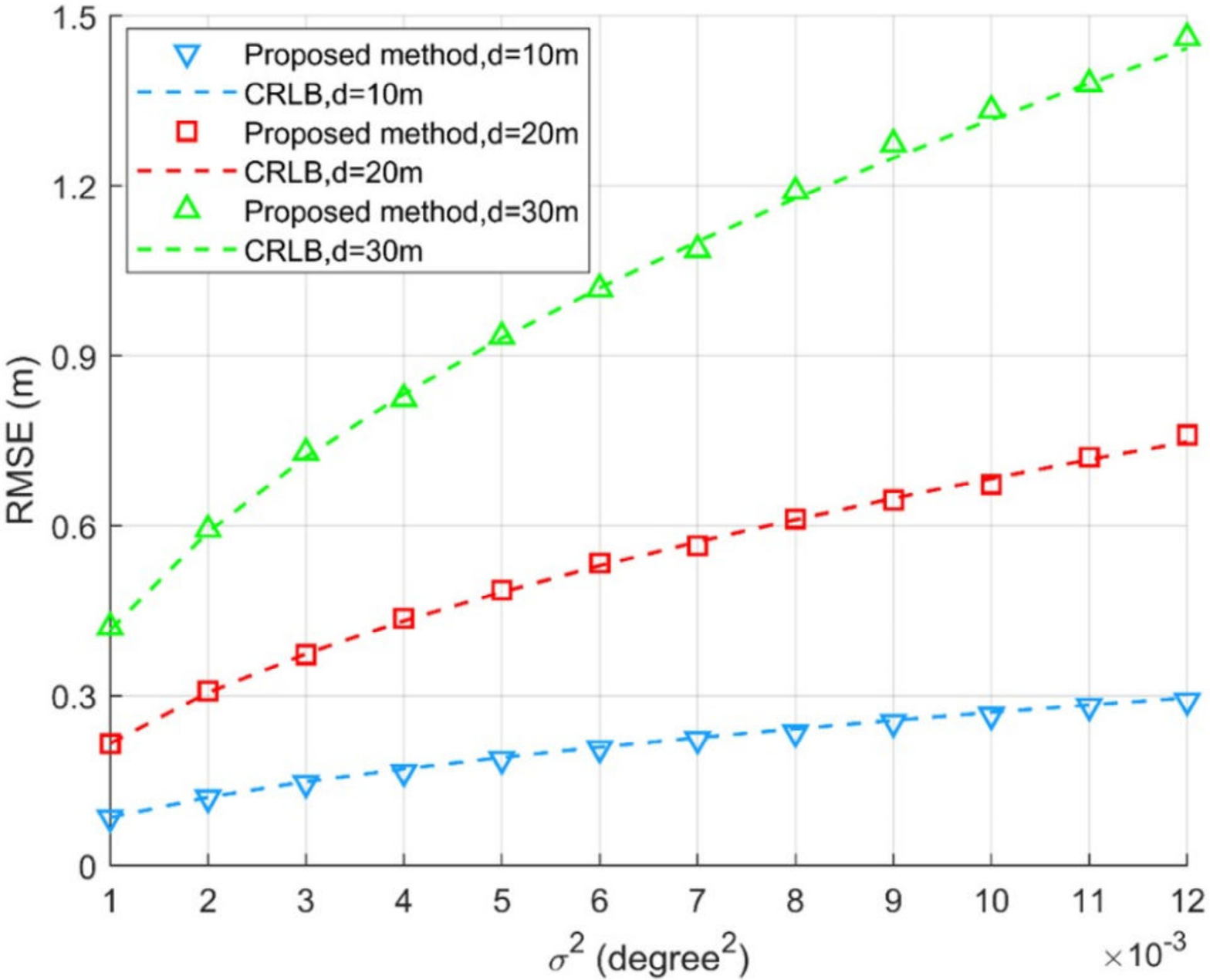}\\
	\caption{RMSE versus variances of estimated angles when the BS is divided into three subarrays spaced 1m apart and the horizontal distance from the target to the BS is 10m, 20m and 30m.}
	\label{fig_2DAOARMSE}
\end{figure}

In Fig. \ref{fig_2DAOARMSE}, the RMSE over the variance of estimated angles is plotted. The root mean square error (RMSE) is adopted in simulations, which is given by 
\begin{equation}
	RMSE = \sqrt{\frac{1}{n}\sum_{n_t=1}^{N_t}\left(\hat{\theta}_{0,n_t}-\theta_0\right)} \nonumber
\end{equation}
where $N_t$  is the number of simulations, which is set as 12000 in this simulation. In this simulation, the array of the BS is divided into three subarrays spaced 1m apart. It can be seen that the proposed method can achieve the CRLB. With the help of the ultra-massive MIMO systems, when the array is divided into more subarrays, the RMSE of the proposed method will lower than 1m. Thus, it is possible to increase the accuracy to decimeter-level or centimeter-level by increasing the number of receive antennas at BS.

\section{Applications and challenging problems}
The DOA estimation will play a more and more important role in the engineering applications, especially in wireless communications. However, some problems still exit. In this section, we will show these promising applications and problems in B5G and 6G communication systems.

In the secure wireless communication systems, it is essential to achieve the accurate channel state information (CSI). In directional modulation (DM) and secure and precise wireless transmission (SPWT) systems, to obtain a high secure rate, artificial noises are generated and projected to the direction without the desired user \cite{hu2016cl,shuSPWT2018jsac,zhou2019jsac,lu2019optimal,hu2017access,shen2019tvt,shuDM2021tcom}. Thus, with the help of artificial noises, only the disturbed confidential messages can be received by the eavesdropper and the signals sent to the user will not be interfered \cite{shu2020net}. However, if the direction of the desired user is inaccurate in DM systems, it is hard to project the artificial noise vector into the null space of the steering vector of the desired user. Then, the signals will be disturbed by the artificial noise, which resulting in the slump of the secure rate. Although some robust methods were proposed in \cite{shu2016access,shu2018sj}, the effect of the imperfect CSI is still great. The DOA estimation using massive MIMO can provide the accurate direction for this system.

Intelligent reflecting surface (IRS) is new technology to improve the performance and reduce the cost for the wireless communication systems \cite{wu2020cm}. IRS is able to change the phase and amplitude of the incident signal and reflect that to the users. It can address many problems, such as the user cannot receive the signal in a dead zone. Different from the relay, IRS is consisted of the low-cost passive equipment reflecting elements. Hence, it is also low power consumption, which is very suitable to the future wireless communication networks. The IRS has been studied in many applications, like the multiuser communication system \cite{wu2019twc}, physical layer security \cite{hong2020tcom,shi2021cc,shuRIS2021tcom}, wireless powered communication networks \cite{shiRIStcom}, Internet-of-Things (IOT) \cite{Basharat2022wc}, indoor mm-wave networks \cite{tan2018infocom} and so on. In \cite{lu2021sl}, the target detection in IRS was considered. All required angles in IRS-aided systems can be acquired by DOA estimation. However, existed methods of DOA estimation in IRS are integration of IRS and traditional methods \cite{chen2021wcl,ma2021twc,li2023ojcs}. How to design low-complexity and high-performance customized methods for IRS is still a open challenging problem.

However, there are still many challenges to be handled in DOA estimation using massive MIMO systems. Several important ones are summarized as follow:

\begin{enumerate}
	\item Huge computational complexity of DOA measurements for massive or ultra-massive MIMO systems. For example, subspace-based methods require the EVD, whose computational complexity is $N^3$  (float-point operations) FLOPs. That brings significantly computational burden, especially when the number of antennas is very large in the future wireless networks. Similarly, how to dramatically reduce the complexity in the estimation of the covariance matrix, EVD and root-finding is also an open problem.
	\item Angle ambiguity is still a challenge for DOA estimation in the HAD structure. Many existing methods need many data-blocks to find out the correct direction, which result in long processing delay. Some methods needing one data-block are based on the cross-correlation, whose performance is worse than subspace-based methods. “Super-resolution” methods that costs only one data block are urgently needed for the HAD structure.
	\item Furthermore, these novel methods are designed for a single emitter, including the methods for the passive target detection, DOA estimation with HAD structure, and AOA localization. Algorithms with super-resolution to deal with multiple co-channel signals are in need.
	\item DOA estimation methods are still incomplete for future applications. To the best of our knowledge, although the exiting methods can be directly used in the low-resolution structure and mixed-ADC structure, there is no specific algorithm. Both structures call for new methods. Furthermore, methods are still scarce for some new systems, like the hybrid UCA and IRS.
	\item In the future wireless communication system, an extremely accuracy location technology is required. Whether a precision of centimeter-level for the location could be realized by a single BS is a challenge. Moreover, if it is practicable, how to reduce the circuit cost and power consumption to make it can be widely used in the future applications? Fortunately, in mm-wave communications, the range of a cellular is very small due to the tremendous signal attenuation. And, the BS is equipped with tens of thousands of antennas. Thus, the high-resolution DOA estimation is available. This pave the way for the centimeter-level source location.
\end{enumerate}

\section{Conclusion}
In this paper, we reviewed the latest developments and presented research results in the DOA estimation using massive receive MIMO. DOA estimation has attracted more and more interest for its applications in wireless communications, IOT, mobile communication, navigation, tracking, passive detection and other fields requiring positioning. It is anticipated that DOA estimation will play an important role, not only due to its ability of precisely positioning, but also for its passive characteristic. DOA estimation and localization using massive MIMO provides high-resolution low-cost solutions for the location-aware system. A new 3D AOA-intersection geometric center localization method was proposed to achieve the CRLB by using a single BS. This method will enable an ultra-high localization performance like centimeter-level for the future applications, such as B5G and 6G. Therefore, we can conclude that the DOA estimation using massive MIMO systems will ultimately benefit the quality of life of people in need.

\ifCLASSOPTIONcaptionsoff
  \newpage
\fi

\bibliographystyle{IEEEtran}
\bibliography{mag_ref}

\end{document}